\documentclass[12pt]{iopart}

\usepackage{amssymb}
\usepackage[pdftex]{color}
\usepackage{graphicx}

\newcommand{\pdag}{{\phantom{\dag}}}

\begin{document}
\title{Dynamics of many-body localization in the presence of particle loss}

\author{EPL~van Nieuwenburg$^{1,2}$, J~Yago Malo$^3$, AJ~Daley$^3$ and MH~Fischer$^1$}
\address{$^1$ Insitute for Theoretical Physics, ETH Zurich, 8093 Zurich, Switzerland}
\address{$^2$ Insitute for Quantum Information and Matter, Caltech, Pasadena, CA 91125, USA}
\address{$^3$ Department of Physics and SUPA, University of Strathclyde, Glasgow G4 0NG, Scottland, UK}

\date{\today}

\begin{abstract}
        At long times, residual couplings to the environment become relevant even in the most isolated experiments, a crucial difficulty for the study of fundamental aspects of many-body dynamics.
	A particular example is many-body localization in a cold-atom setting, where incoherent photon scattering introduces both dephasing and particle loss.
	Whereas dephasing has been studied in detail and is known to destroy localization already on the level of non-interacting particles, the 
	effect of particle loss is less well understood.
	A difficulty arises due to the `non-local' nature of the loss process, complicating standard numerical tools using matrix product decomposition. 
	Utilizing symmetries of the Lindbladian dynamics, we investigate the particle loss on both the dynamics of observables, as well as the structure of the density matrix and the individual states.
	We find that particle loss in the presence of interactions leads to dissipation and a strong suppression of the (operator space) entanglement entropy.
	Our approach allows for the study of the interplay of dephasing and loss for pure and mixed initial states to long times, which is important for future experiments using controlled coupling of the environment. 
\end{abstract}



\section{Introduction}
Advances in synthetic quantum systems allow for the simulation of paradigmatic model Hamiltonians in highly controlled settings, closely corresponding to ideal closed systems. 
As a residual coupling to the environment persists even in these systems, it is crucial to address the effects of such non-zero coupling, in particular when studying properties connected to slow intrinsic timescales. 
Such simulations are well established in the context of quantum-optics~\cite{carmichael:1999}, where dissipative effects of the environment can be captured in well-controlled microscopic models~\cite{daley:2014, mueller:2012}. 
Recently, experiments with cold atoms in optical lattices have demonstrated controlled coupling to an environment, through incoherent light scattering and particle loss~\cite{lueschen:2017}. Thus, it is important to have a means to benchmark and better understand these processes through numerical simulations.
A renewed interest for this problem for the case of many-body systems has focused mostly on bosonic or spin systems, but the efficient simulation of dissipative dynamics for fermionic many-body systems has only started to attract attention~\cite{sarkar:2014}. A particular problem arises for fermionic systems due to their statistics, introducing signs whenever particles are exchanged with the environment. When working in an occupation-number basis, we can formally write these signs as the appearance of string operators keeping track of the operator commutations to reach a given site, which renders the formally short-ranged problem long ranged.

Many-body localization presents an example of a model system that has attracted considerable attention in recent years~\cite{nandkishore:2015, altman:2015}. Isolated many-body-localized systems fail to thermalize even at infinite time scales, providing a novel paradigm contrasting ergodic behavior.
Recently, such systems have been studied in trapped-ion settings~\cite{smith:2016} and with cold atoms in optical lattices~\cite{schreiber:2015, bordia:2016}.
However, in the latter both particle loss and dephasing due to incoherent light scattering are naturally present~\cite{lueschen:2017} and destroy localization at long times.
Several theoretical works have studied the influence of a dephasing bath on the relaxation dynamics in a many-body-localized system in the Markov approximation~\cite{medvedyeva:2016,fischer:2016c,  levi:2016}, which is well suited for typical experimental parameters. The effect of particle loss, however, has only been addressed on a phenomenological level providing insight for special initial states~\cite{fischer:2016c}. 

The time evolution of a system coupled to a Markovian bath is governed by the Lindblad equation for the density matrix
\begin{equation}
    \frac{\partial}{\partial t}{\rho} = \mathcal{L}[\rho] = -i[H, \rho] + \sum_{i}\gamma_i \Big(L^\pdag_{i}\rho L_{i}^\dag - \frac12\{L_{i}^\dag L_{i}^\pdag, \rho\}\Big),
  \label{eq:lindblad}
\end{equation}
where $\mathcal{L}$ is referred to as the Lindbladian operator and in the last equality the first term describes the unitary time evolution with Hamiltonian $H$ and the second term the coupling to the environment.
The system operators $L_i$ are called jump operators and directly couple to the bath with rate $\gamma_i$.
We will focus on the case provided by spinless fermions on a one-dimensional lattice with random on-site energies and nearest-neighbor interactions, 
\begin{equation}
  H = -J\sum_i (c^{\dag}_i c^{\phantom{\dag}}_{i+1} + h.c.) + \sum_i V_i n_i + U\sum_{i} n_in_{i+1}.
  \label{eq:HIntAnderson}
\end{equation}
Here, $n_i = c_i^{\dag}c^\pdag_i$, where $c^\dagger_i$ creates a fermion on site $i$, and $V_i \in [-h, h]$ are on-site energies, which are independent random variables. This Hamiltonian has been extensively studied as a model (closed) system and (for $U=2J$) is expected to be localized for $h\gtrsim 7.2 J$~\cite{pal:2010, luitz:2015, serbyn:2015}.
We are then interested in processes that occur naturally in experiments, namely particle loss, i.e., $L_i = c_i$, as well as dephasing or local density measurement with $L_i = n_i$~\cite{pichler:2010}. 

A powerful approach for the simulation of both closed and open one-dimensional systems is the matrix product decomposition of states and operators~\cite{schollwock:2011}. In general, a state can be written as
\begin{equation}
    |\psi\rangle = \sum_{i_1, i_{2}, \dots, i_L}\Tr(A^{i_1}\cdots A^{i_L}) |i_1 \cdots i_L\rangle,
    \label{eq:mps}
\end{equation}
where each $A^{i_n}$ is a matrix of maximal dimension $\chi\times\chi$ per physical index $i_n$.
For spinless fermions, $i_n$ corresponds to the local occupation, i.e., the Hilbert space is spanned by all $|n_1 n_2\dots\rangle = \Pi_{\{i\}} c^\dag_i|\Omega\rangle$, where the product runs over the set $\{i\}$ of occupied sites and $|\Omega\rangle$ represents the vacuum state with no particles.
Equivalently, a general operator $\mathcal{O}$ can be represented in matrix product form~\cite{verstraete:2004}. For even combinations of single-particle operators with a finite support between sites $i$ and $i+l$, such as given in short-ranged Hamiltonians, these operators can be written as 
\begin{equation}
    \mathcal{O} = \mathbb{I}\times \mathbb{I}\times\cdots \mathcal{O}_{i, i+l} \cdots \times \mathbb{I},
    \label{eq:ops}
\end{equation} 
where $\mathbb{I}$ denotes a local identity operator and the operator $\mathcal{O}_{i, i+l}$ only acts on the physical indices of sites $i$ to $i+l$. With only local updates required, this allows for the efficient computation of time evolution through time-evolving block decimation (TEBD)~\cite{vidal:2004}. 

This locality is, however, lost, if we want to annihilate (create) a single particle from a site $i$, since the fermionic statistics requires the introduction of a non-local string of phases (c.f. the Jordan-Wigner transformation).  
In the following, we show how using an appropriate symmetry, the `bond parity', a matrix-product-operator (MPO) approach for the density matrix allows for the simulation of the relaxation dynamics of an MBL system coupled to a Markovian bath. 
Simulating the full density matrix simplifies the implementation of mixed initial states and further gives us access to longer times in the presence of a bath. We investigate this latter point by calculating the operator-space entanglement entropy (OSEE), which measures the factorizability of the density matrix~\cite{prosen:2007} and is a measure for the efficiency of the matrix product formalism. While this entropy shows a logarithmic growth for both closed systems and dephasing bath, thereby resembling the entanglement entropy of the pure-state evolution, particle loss leads to a rapid decrease.
Finally, we employ a quantum trajectory approach as an alternative view on the entanglement generated during the dynamics and to compare computational efficiency.

\section{Methods}
\subsection{Matrix-Product-Operator Formalism}
In general, we can write the density matrix as
\begin{equation}
    \rho = \sum_{n_1, n_2,\cdots}\sum_{n_1', n'_2, \cdots} \rho^{\{n_i\}}_{\{n'_i\}} |n_1\cdots\rangle\langle\cdots n'_1|.
    \label{eq:densitymatrix}
\end{equation}
In this basis, the term $L_i^{\phantom{\dag}}\rho L_i^\dag$ with jump operators $L_i= c_i$, which describe the loss of particles, will acquire a sign depending on the number of particles to the left of site $i$ both in the bra and the ket element, 
\begin{equation}
    c_i |\cdot n_i\;\cdot\;\rangle\langle\;\cdot\; n'_i\cdot| c_i^\dag =
    (-1)^{N_{<i} + N'_{<i}}|\cdot\hat{n}_i\;\cdot\;\rangle\langle\; \cdot\; \hat{n}'_i \cdot|,
    \label{eq:loss}
\end{equation}
where $N_{<i} = \sum_{j<i} n_j$ and the state on the right side is equal to the state on the left side apart from particles at site $i$ missing. The sign of the (local) term thus depends on the total sum of particles up to site $i$, which we will refer to as bond parity $\mathcal{P}_i = (-1)^{N_{<i} + N_{<i}'}$. This dependence on the particle number renders the operators non-local, and standard TEBD approaches can not be used anymore. Note that for all the other terms in (\ref{eq:lindblad}), there is no additional sign, as the operators always come in pairs, thus canceling out the strings.

\begin{figure}[bt]
    \centering
    \includegraphics[width=0.7\textwidth]{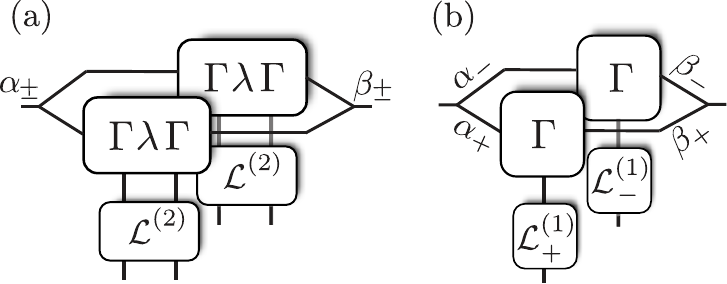}
    \caption{Schematic of a TEBD step for updating the MPO matrices $\Gamma$ and $\lambda$, preserving the subspaces $\mathcal{P}=\pm1$. (a) For the bond terms, the same operator $\mathcal{L}^{(2)}$ is applied in each subspace. For the case (b) of the single-site Lindblad superoperators, each block $\mathcal{P}=\pm1$ has its own single-site operator $\mathcal{L}^{(1)}_{\pm}$ applied in the respective subspace. The sets of indices $\alpha_{\pm}$ and $\beta_{\pm}$ correspond to the subspaces with associated quantum numbers $\mathcal{P}=\pm1$.}
    \label{fig:method}
\end{figure}

In order to deal with the additional string or sign associated with the loss operators, we first separate the full Lindbladian into bond and single-site terms
\begin{equation}
    \mathcal{L}[\rho] = \mathcal{L}^{(2)}[\rho] + \mathcal{L}^{(1)}[\rho],
    \label{eq:splitL}
\end{equation}
where $\mathcal{L}^{(2)}$ contains the unitary evolution, and $\mathcal{L}^{(1)}$ 
contains only single-site terms.
Only these latter terms can suffer from a relative sign, such that for particle loss, i.e., $L_i = c_i$, we split it further into
\begin{equation}
    \mathcal{L}^{(1)}_{i, \pm}[\rho] = \gamma_i\Big(\pm c^{\phantom{\dag}}_i \rho c_i^\dag  - \frac12\{c_i^\dag c_i^{\phantom{\dag}}, \rho\}\Big),
    \label{eq:even_odd}
\end{equation}
acting only locally.
Importantly, while the loss term changes the number of particles, both single-site and bond terms preserve the bond parity on the left and right bond from where they act and hence, all updates can be performed in the subspaces of $\mathcal{P}_i = \pm 1$. 

We have implemented the time evolution within the superoperator generalization of TEBD for density matrices~\cite{zwolak:2004}, where the density matrix (\ref{eq:densitymatrix}) is written as
\begin{equation}
    \rho = \sum_{j_1,j_2,\dots } \Tr[\lambda^0 \Gamma^{j_1}\lambda^1\cdots \Gamma^{j_L}\lambda^L] \sigma^{j_1}\otimes\cdots\otimes\sigma^{j_L}.
    \label{eq:mp0}
\end{equation}
Here, $\sigma^{j}$ is a basis for the $2\times2$ matrices, e.g., the Pauli matrices together with the identity $\sigma^0$. 
The physical indices then come with respective (bond parity) quantum numbers $+1$ for the diagonal $\sigma^0$ and $\sigma^3$ and $-1$ for $\sigma^1$ and $\sigma^2$.
In (\ref{eq:mp0}), we follow the notation of Vidal~\cite{vidal:2003}, where the singular values are stored in diagonal matrices $\lambda^i$.
Finally, we separate the bond operators into even and odd bond operators $\mathcal{L}^{(e)}$ and $\mathcal{L}^{(o)}$ and use a Trotter decomposition for the time-evolving block decimation,  
\begin{equation}
	e^{\mathcal{L} t }\approx \Big(\Pi e^{\frac{\Delta t}{2N}\mathcal{L}^{(1)} }\Pi e^{\frac{\Delta t}{2N}\mathcal{L}^{(e)} }\Pi e^{\frac{\Delta t}{N}\mathcal{L}^{(o)}}\Pi e^{\frac{\Delta t}{2N}\mathcal{L}^{(e)}}\Pi e^{\frac{\Delta t}{2N}\mathcal{L}^{(1)} }\Big)^N.
    \label{eq:Tevolve}
\end{equation}
Similarly to other implementations for conserved quantities, we keep track of the parity by enriching each bond index $\alpha$ with  a map $\mathcal{P}:\alpha\rightarrow \pm 1$. The bond operators are applied as shown in figure~\ref{fig:method}(a) for updating the MPO matrices represented by $\Gamma$ and $\lambda$. 
However, unlike standard TEBD implementations respecting quantum numbers~\cite{muth:2011}, the correct operator has to be chosen for the single-site term in order to preserve the parity subspaces, see figure~\ref{fig:method}(b).
Note that while one could further simplify the calculation by working with occupation quantum numbers for unitary evolution and dephasing, this is no longer true when considering particle loss.

\section{Results}

We have employed the above formalism to calculate the effect of a general bath including both particle loss and dephasing (with respective coupling rates $\gamma_l$ and $\gamma_d$) on a system described by (\ref{eq:lindblad}) and (\ref{eq:HIntAnderson}). 
Following previous literature, we calculate the time evolution of an initial occupation imbalance, $\mathcal{I}(t) = [n_e(t) - n_o(t)]/[n_e(t) + n_o(t)]$ with $n_{e/o}(t)$ the density of particles on even/odd sites to access the system's relaxation dynamics.
Unless noted otherwise, we start from a pure state given by the perfect density wave, i.e., $|\psi_0\rangle = |0101\dots\rangle$ and use numerical parameters $\chi=100$ (bond dimension) and $\Delta t=0.1\; [1/J]$.

\begin{figure}[bt]
    \centering
\includegraphics[width=\textwidth]{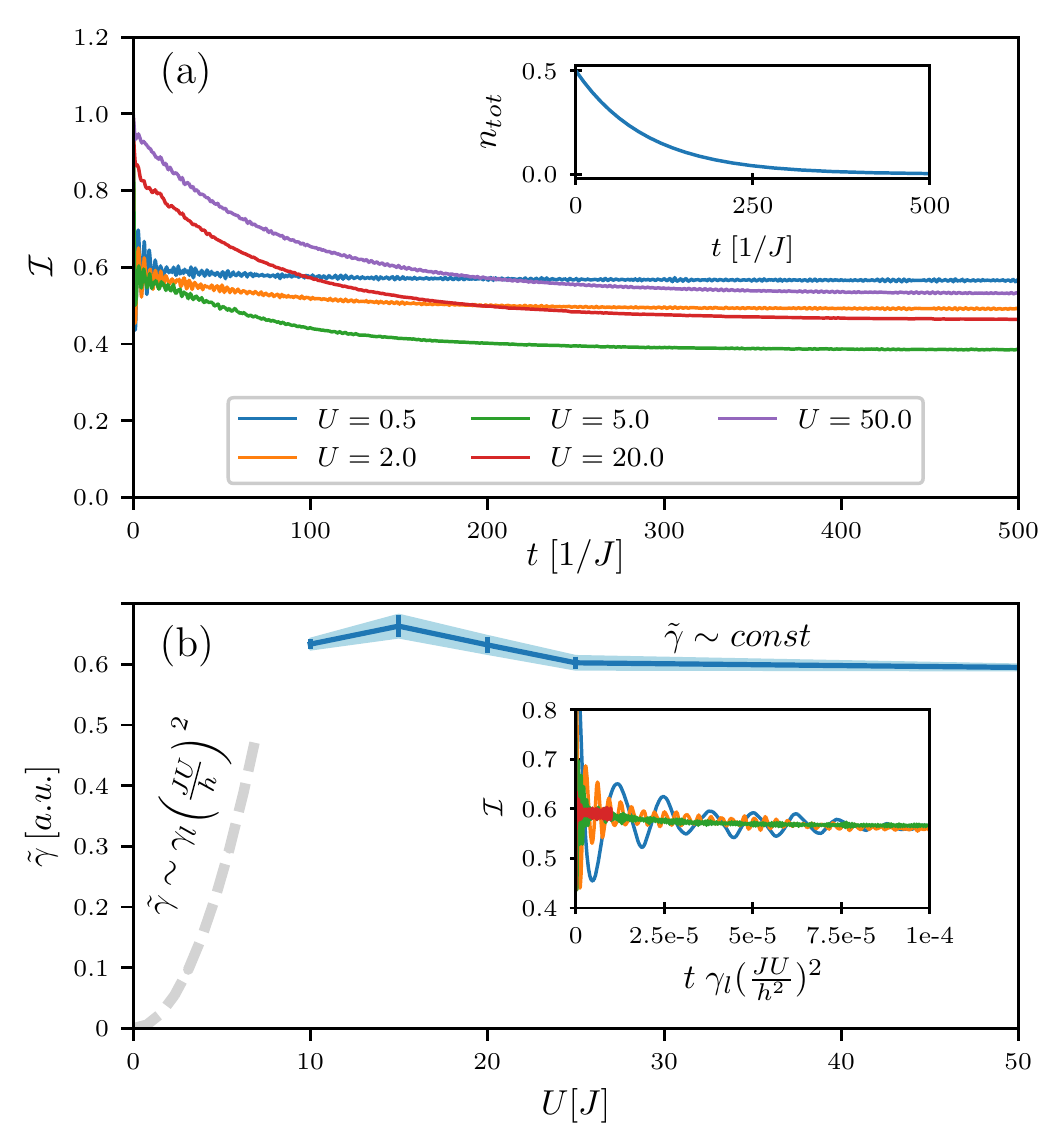}
\caption{(a) Dynamics of the imbalance and particle density (inset) for loss only ($\gamma_l=0.01$) for $h=10J$, $N=20$, and $U/J=0.5\ldots 50.0$. (b) Effective decay rate of the imbalance extracted via a fit to (\ref{eq:It}) for $U\geq10J$. For $U/J = 0.5, 1.0, 2.0$ and $5.0$, the qualitative behavior (gray dashed line) is confirmed via a collapse of the imbalance time traces by scaling the time with $\gamma_l J (J U/h^2)^2$ (inset).}
    \label{fig:loss}
\end{figure}

\subsection{Pure loss}
The case describing a bath that induces no dephasing but only loss can be understood in both the weak- and strong-interaction limit from a phenomenological perspective based on integrals of motion~\cite{fischer:2016c}, and thus provides a good benchmark to study first.
In the absence of interactions loss does not change the normalized imbalance since the occupancy on even and odd sites follows the same exponential decay. 
However, upon introducing a nearest-neighbor interaction $U\neq 0$ the imbalance is expected to decrease. 
This is due to the change of the local potential a particle experiences as a neighboring particle is removed, which may enable hopping.
Only after the particle density has been reduced sufficiently does the imbalance stabilize as the system becomes effectively non-interacting.

This behavior can be qualitatively described in a mean-field picture, where the time evolution of the imbalance can be approximated by~\cite{fischer:2016c, levi:2016}
\begin{equation}
    \frac{d\mathcal{I}}{dt} \approx -\gamma_{\rm eff}(t) \mathcal{I}(t) \approx -\gamma n(t) \mathcal{I}(t) = -\tilde{\gamma} e^{-\gamma_l t} \mathcal{I}(t),
    \label{eq:diffI}
\end{equation}
where $n(t) = n_0 \exp(-\gamma_l t)$ is the particle density and $\tilde{\gamma} = \gamma n_0$ is an effective decay rate. Integrating (\ref{eq:diffI}) yields the time dependence for the imbalance
\begin{equation}
    \mathcal{I}(t) \propto \exp\Big\{-\Big[\frac{\tilde{\gamma}}{\gamma_l} (1-e^{-\gamma_l t})\Big]^\beta\Big\}.
    \label{eq:It}
\end{equation}
Here, we have introduced an exponent $\beta$ phenomenologically to account for the fact that the mean-field picture is a simplification~\cite{fischer:2016c} and to yield better agreement with the numerical results.

Figure~\ref{fig:loss}(a) shows the imbalance relaxation due to loss for a system of length $N=20$ sites, $h=10J$, and various interactions. As expected, the imbalance relaxes initially with a rate increasing with the interaction strength before saturating once the density has dropped sufficiently (see inset). For small interactions $U\ll h$, the initial decay can be calculated perturbatively and is expected to scale as $\gamma_l (J U/h^2)^2$~\cite{fischer:2016c}. For large interactions, on the other hand, the initial decay becomes independent of the interaction strength. This dependence of the effective decay rate $\tilde{\gamma}$ on the interaction strength is shown in figure~\ref{fig:loss}(b). Equation~(\ref{eq:It}) yields a robust fit to the numerical time traces with $\beta\approx0.9$ for all interactions $U\geq10J$ and allows for a direct extraction of the decay rate. For comparison, the error bars denote the difference when fitting to a fixed $\beta=1$. Unfortunately, the fitting does not work as robustly for smaller interactions and instead, we confirm the initial decay rate $\tilde{\gamma}\sim U^2$ via a data collapse of the time traces shown in the inset of figure~\ref{fig:loss}(b). 


\subsection{Dephasing and loss: weak interactions}
\begin{figure}[tb]
    \centering
    \includegraphics[width=\textwidth]{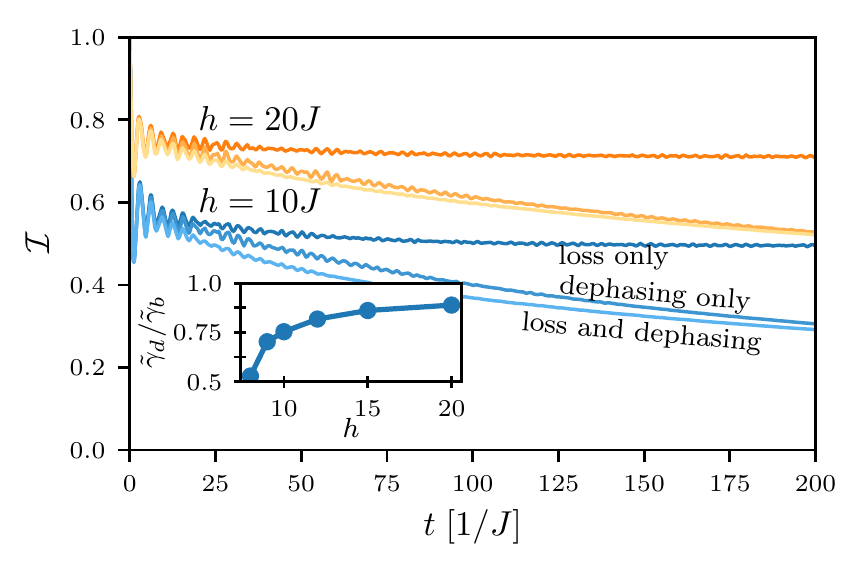}
    \caption{Imbalance dynamics in the presence of dephasing and loss for intermediate interactions $U=2J$ and disorder $h=10J$ (lower traces) and $h=20J$. The more rapidly decaying traces represent the cases of both dephasing and loss combined [$(\gamma_d, \gamma_l) = (0.02, 0.02)$]. The `loss only' (`dephasing only') traces have $(\gamma_d, \gamma_l)=(0.0, 0.02)$ [$(0.02, 0.0)$]. The inset shows the ratio $\tilde{\gamma}_d/\tilde{\gamma}_b$ of the decay rates extracted from pure dephasing and the combination of both dephasing and loss, respectively. For larger disorder the decay rate is dominated by the dephasing, as shown by this ratio approaching unity.}
    \label{fig:U2AllCDW}
\end{figure}

Having established the dynamics of the system when only particle loss is present, we next address the interplay of both dephasing and loss. As can be seen in figure~\ref{fig:U2AllCDW} the inclusion of dephasing processes for weak to intermediate interactions has a detrimental effect on the localization. In this regime, the dephasing processes directly lead to dissipation, and additional particle loss provides only a subleading effect. Regardless, the relative effect of loss is largest close to the critical disorder strength and decreases for stronger disorder. This is in agreement with recent experiments with spinful fermions~\cite{lueschen:2017}. The inset of figure~\ref{fig:U2AllCDW} emphasizes this trend by comparing the effective initial decay rate for dephasing only with dephasing and loss combined, extracted using a stretched exponential decay for both time traces for various disorder strengths.

\subsection{Dephasing and loss: strong interactions}
\begin{figure}[tb]
    \centering
    \includegraphics[width=\textwidth]{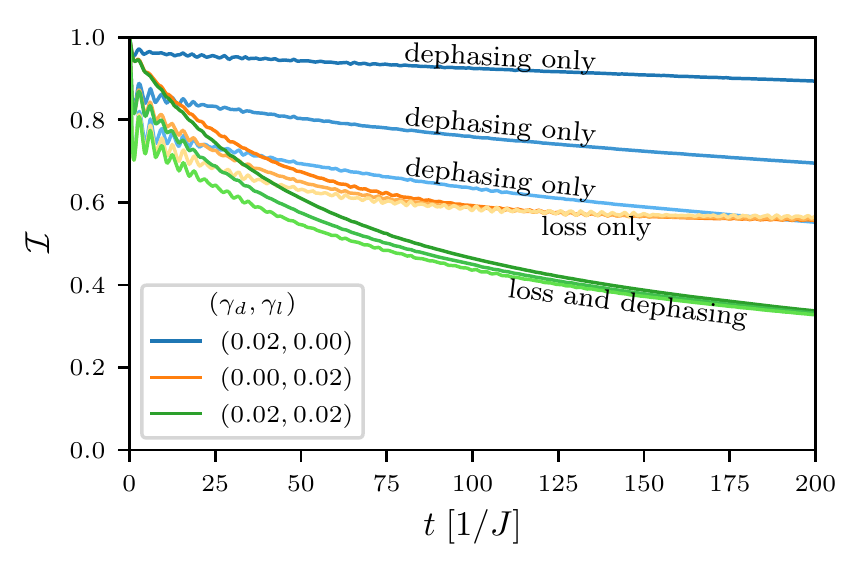}
    \caption{Imbalance evolution for for pure dephasing (blue), pure loss (orange), and both (green) for densities $n=0.5, 0.4$, and $n=0.3$ (top to bottom, increasingly lighter colors) for large interactions, $U=50J$. Here, $\Delta t = 0.025$ $[1/J]$ was used.}
    \label{fig:U50all}
\end{figure}

As the interaction strength increases, the initial density-wave state at half filling has more and more overlap with an eigenstate of the system. In such a situation, the dephasing-induced hopping of particles is suppressed as $1/U^2$ and dephasing has vanishing effect. Hence, the particle-loss term is expected to dominate, which is clearly shown in figure~\ref{fig:U50all} for $U=50J\gg h$ comparing the dynamics of loss only, dephasing only, and a combination of the two. Only after sufficient particles have been lost does the dephasing become relevant, leading to additional diffusion. 

For a more general investigation of the dynamics at large interactions, it is thus necessary to not only focus on the perfect density wave with $n=0.5$ as the initial state, but also on density-wave states with lower overall density. The matrix product operator approach is well suited for implementing these states, and figure~\ref{fig:U50all} additionally shows the imbalance decay for pure dephasing, pure loss, and the combination for additional densities $n=0.3, 0.4$. For these densities, the initial states have less overlap with a single eigenstate and the imbalance drops to a lower value at short times correspondingly. In addition, the influence of dephasing increases since for the initial state not every second site is occupied and thus the effect of the nearest-neighbor interaction is decreased. In the limit of vanishing initial density, we would thus recover the non-interacting result, where loss has no effect on the imbalance dynamics.
\subsection{Evolution of entanglement entropy}
To investigate the efficiency of our approach we consider the density matrix's entanglement with respect to the middle bond of the state with $n=0.5$. The OSEE provided by the MPO representation of the density matrix~(\ref{eq:mp0}), i.e., $S_{\sharp} = -2 \sum_{\alpha} (\lambda^{i}_{\alpha})^{2}\log \lambda^{i}_{\alpha}$ for bond $i=N/2$ is a measure of the factorizability of the density matrix and thus its unbound growth signals the breakdown of the MPO approach.

\begin{figure}[tb]
    \centering
    \includegraphics[width=\textwidth]{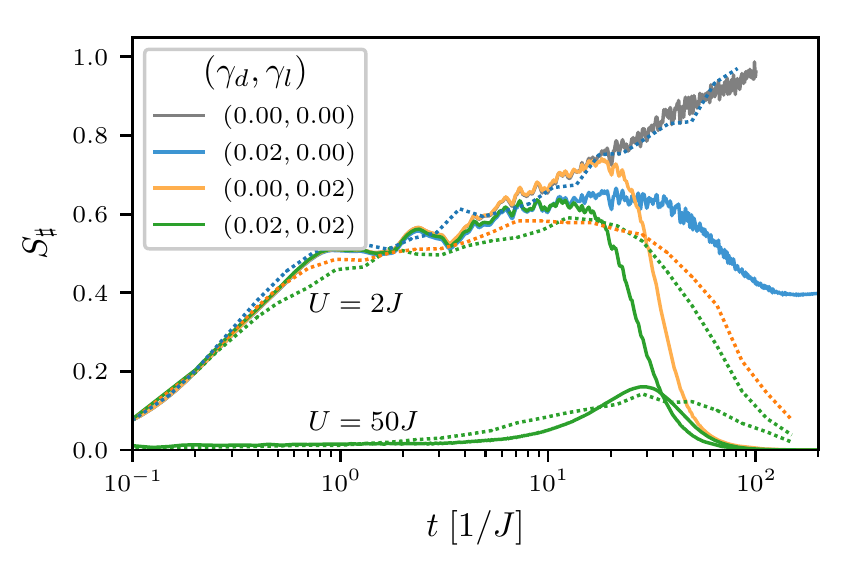}
    \caption{Evolution of the OSEE for $U=2J$ and $U=50J$ and disorder strength $h=10J$ (solid lines) and the entanglement entropy obtained from quantum trajectory simulations (dashed lines). We used a base $d$, corresponding to the local Hilbert space dimension in each method, for the logarithm of the respective entropy for better comparison. The timestep used for the OSEE was $\Delta t = 0.005~[1/J]$. For the quantum trajectories a timestep of size $\Delta t = 0.001~[1/J]$ and bond dimension $\chi = 250$ were used. 
}
    \label{fig:U2OSEE}
\end{figure}

Figure~\ref{fig:U2OSEE} compares the OSEE evolution for various baths starting from the perfect density wave, a product state. For $U=50J$, the OSEE stays close to zero until particle loss removes the system from being close to an eigenstate, confirming the observations of the previous section for $n=0.5$. For $U=2J$, the closed system OSEE shows a $\log t$ growth, in accordance with the (pure state) entanglement entropy, which limits the accessible times to $t\approx 100/J$. Pure dephasing leads to a decrease of the OSEE at intermediate times, but a characteristic $\log t$ growth sets in at long times~\cite{medvedyeva:2016}. With the inclusion of loss, however, the OSEE quickly drops to zero after times $t \sim 1/\gamma_l$ irrespective of interaction strength, since at this timescale the loss of particles becomes significant and effectively prevents long-range correlations. Note, however, that there is still a significant number of particles left in the system at this time.

Finally, we compare the evolution of the OSEE with the entanglement entropy obtained via a quantum trajectory method~\cite{daley:2014, yagomalo:2017tmp} as shown with dashed lines in figure~\ref{fig:U2OSEE}. For comparison, we calculated the entanglement entropy using a logarithm with base equal to the local Hilbert space dimension for both the quantum trajectory method ($d=2$) and the density matrix evolution ($d=4$). While the overall behavior of the two entropies is similar, the entropy from the trajectory method shows both a slower growth and a longer decay. This difference can be understood as follows: First, for a trajectory the local density measurement (representing the dephasing) does not affect the entanglement by resetting it to a product state. This behaviour is demonstrated by the blue dashed line in figure~\ref{fig:U2OSEE}. Second, the probabilistic nature of the quantum trajectory approach includes trajectories where jumps, i.e., coupling to the bath, occur very early, or not until late in the time evolution. This resembles more closely the actual experimental situation, where only averaging over many runs results in the statistical behavior described by the density matrix. 
However, for disordered systems, where the initial entropy growth is slow, our MPO approach provides a more efficient method for the calculation of the dynamics induced by a Markovian bath. 

\section{Conclusion}
We have implemented the Lindbladian time evolution for a fermionic system coupled to a bath through dephasing and loss employing a matrix-product-operator formalism. For this purpose, we have identified the `bond parity' as a symmetry of the Lindbladian evolution and implemented this symmetry in a quantum-number-conserving TEBD calculation. Our approach allows for the efficient calculation of the time evolution of any (mixed) initial state in the presence of a bath to long times and thus for the comparison to experimental setups that have recently become available.

As a specific example of interest, we have used our approach to investigate the effects of loss on the dynamics of an otherwise many-body-localized system%
. We have confirmed the behavior of an initial charge imbalance expected from a phenomenological model of pure loss in the limits of weak and strong interaction. Importantly, we have shown how dephasing and loss have different significance for inducing diffusion at weak compared to strong coupling, as well as close to the localization transition compared to deep in the localized phase.

The interplay between interactions, loss, and dephasing that we study here could be realised in ongoing experiments. Recent advances in cooling and observing atoms that have a large intrinsic dipole-dipole interaction such as Erbium allow for the realisation of lattice models with significant off-site interactions~\cite{baier:2016}. Beyond these specific systems, our matrix-product-operator implementation that retains parity information provides a new general tool for simulating particle loss and other dissipative processes where we have to account for such signs, significantly expanding the available possibilities for efficient simulation of dynamics induced by a Markovian bath.

\ack
EvN gratefully acknowledges funding by the Swiss National Science Foundation through grant P2EZP2 172185. Work at Strathclyde is supported in part by the US Air Force Office of Scientific Research grant number FA2386-14-1-5003.

\section*{References}
\bibliographystyle{unsrt}

\begin{thebibliography}{10}

\bibitem{carmichael:1999}
Howard~J Carmichael.
\newblock {\em {Dissipation in Quantum Mechanics: The Master Equation
  Approach}}, pages 1--28.
\newblock Springer Berlin Heidelberg, Berlin, Heidelberg, 1999.

\bibitem{daley:2014}
Andrew~J Daley.
\newblock {Quantum trajectories and open many-body quantum systems}.
\newblock {\em Advances in Physics}, 63(2):77--149, 2014.

\bibitem{mueller:2012}
Markus M{\"{u}}ller, Sebastian Diehl, Guido Pupillo, and Peter Zoller.
\newblock {Engineered Open Systems and Quantum Simulations with Atoms and
  Ions}.
\newblock {\em Adv. At. Mol. Opt. Phys}, 61(1):1--80, 2012.

\bibitem{lueschen:2017}
Henrik~P L{\"{u}}schen, Pranjal Bordia, Sean~S Hodgman, Michael Schreiber,
  Saubhik Sarkar, Andrew~J Daley, Mark~H Fischer, Ehud Altman, Immanuel Bloch,
  and Ulrich Schneider.
\newblock {Signatures of Many-Body Localization in a Controlled Open Quantum
  System}.
\newblock {\em Phys. Rev. X}, 7:11034, mar 2017.

\bibitem{sarkar:2014}
S~Sarkar, S~Langer, J~Schachenmayer, and A~J Daley.
\newblock {Light scattering and dissipative dynamics of many fermionic atoms in
  an optical lattice}.
\newblock {\em Phys. Rev. A}, 90:23618, aug 2014.

\bibitem{nandkishore:2015}
Rahul Nandkishore and David~A. Huse.
\newblock Many-body localization and thermalization in quantum statistical
  mechanics.
\newblock {\em Annual Review of Condensed Matter Physics}, 6(1):15--38, 2015.

\bibitem{altman:2015}
Ehud Altman and Ronen Vosk.
\newblock {Universal Dynamics and Renormalization in Many-Body-Localized
  Systems}.
\newblock {\em Annual Review of Condensed Matter Physics}, 6(1):383--409, 2015.

\bibitem{smith:2016}
J~Smith, A~Lee, P~Richerme, B~Neyenhuis, P~W Hess, P~Hauke, M~Heyl, D~A Huse,
  and C~Monroe.
\newblock {Many-body localization in a quantum simulator with programmable
  random disorder}.
\newblock {\em Nat Phys}, 12(10):907--911, 2016.

\bibitem{schreiber:2015}
Michael Schreiber, Sean~S Hodgman, Pranjal Bordia, Henrik~P L{\"{u}}schen,
  Mark~H Fischer, Ronen Vosk, Ehud Altman, Ulrich Schneider, and Immanuel
  Bloch.
\newblock {Observation of many-body localization of interacting fermions in a
  quasirandom optical lattice}.
\newblock {\em Science}, 349(6250):842--845, 2015.

\bibitem{bordia:2016}
Pranjal Bordia, Henrik~P. L\"uschen, Sean~S. Hodgman, Michael Schreiber,
  Immanuel Bloch, and Ulrich Schneider.
\newblock Coupling identical one-dimensional many-body localized systems.
\newblock {\em Phys. Rev. Lett.}, 116:140401, Apr 2016.

\bibitem{medvedyeva:2016}
Mariya~V Medvedyeva, Toma{\v{z}} Prosen, and Marko {\v{Z}}nidari{\v{c}}.
\newblock {Influence of dephasing on many-body localization}.
\newblock {\em Phys. Rev. B}, 93:94205, mar 2016.

\bibitem{fischer:2016c}
Mark~H Fischer, Mykola Maksymenko, and Ehud Altman.
\newblock {Dynamics of a Many-Body-Localized System Coupled to a Bath}.
\newblock {\em Phys. Rev. Lett.}, 116:160401, apr 2016.

\bibitem{levi:2016}
Emanuele Levi, Markus Heyl, Igor Lesanovsky, and Juan~P Garrahan.
\newblock {Robustness of Many-Body Localization in the Presence of
  Dissipation}.
\newblock {\em Phys. Rev. Lett.}, 116:237203, jun 2016.

\bibitem{pal:2010}
Arijeet Pal and David~A Huse.
\newblock {Many-body localization phase transition}.
\newblock {\em Phys. Rev. B}, 82:174411, nov 2010.

\bibitem{luitz:2015}
David~J. Luitz, Nicolas Laflorencie, and Fabien Alet.
\newblock Many-body localization edge in the random-field heisenberg chain.
\newblock {\em Phys. Rev. B}, 91:081103, Feb 2015.

\bibitem{serbyn:2015}
Maksym Serbyn, Z~Papi{\'{c}}, and Dmitry~A Abanin.
\newblock {Criterion for Many-Body Localization-Delocalization Phase
  Transition}.
\newblock {\em Phys. Rev. X}, 5:41047, dec 2015.

\bibitem{pichler:2010}
H~Pichler, A~J Daley, and P~Zoller.
\newblock {Nonequilibrium dynamics of bosonic atoms in optical lattices:
  Decoherence of many-body states due to spontaneous emission}.
\newblock {\em Phys. Rev. A}, 82(6):63605, dec 2010.

\bibitem{schollwock:2011}
Ulrich Schollw{\"{o}}ck.
\newblock {The density-matrix renormalization group in the age of matrix
  product states}.
\newblock {\em Annals of Physics}, 326(1):96--192, 2011.

\bibitem{verstraete:2004}
F~Verstraete, J~J Garcia-Ripoll, and J~I Cirac.
\newblock {Matrix Product Density Operators: Simulation of Finite-Temperature
  and Dissipative Systems}.
\newblock {\em Phys. Rev. Lett.}, 93(20):207204, nov 2004.

\bibitem{vidal:2004}
Guifr{\'{e}} Vidal.
\newblock {Efficient Simulation of One-Dimensional Quantum Many-Body Systems}.
\newblock {\em Phys. Rev. Lett.}, 93:40502, jul 2004.

\bibitem{prosen:2007}
Tomaz Prosen and Iztok Pi{\v{z}}orn.
\newblock {Operator space entanglement entropy in a transverse Ising chain}.
\newblock {\em Phys. Rev. A}, 76:32316, sep 2007.

\bibitem{zwolak:2004}
Michael Zwolak and Guifr{\'{e}} Vidal.
\newblock {Mixed-State Dynamics in One-Dimensional Quantum Lattice Systems: A
  Time-Dependent Superoperator Renormalization Algorithm}.
\newblock {\em Phys. Rev. Lett.}, 93:207205, nov 2004.

\bibitem{vidal:2003}
Guifr{\'{e}} Vidal.
\newblock {Efficient Classical Simulation of Slightly Entangled Quantum
  Computations}.
\newblock {\em Phys. Rev. Lett.}, 91:147902, oct 2003.

\bibitem{muth:2011}
Dominik Muth.
\newblock {Particle number conservation in quantum many-body simulations with
  matrix product operators}.
\newblock {\em Journal of Statistical Mechanics: Theory and Experiment},
  2011(11):P11020, 2011.

\bibitem{yagomalo:2017tmp}
J~Yago~Malo, EPL van Nieuwenburg, MH~Fischer, and AJ~Daley.
\newblock {\em in preparation}.

\bibitem{baier:2016}
S.~Baier, M.~J. Mark, D.~Petter, K.~Aikawa, L.~Chomaz, Z.~Cai, M.~Baranov,
  P.~Zoller, and F.~Ferlaino.
\newblock Extended bose-hubbard models with ultracold magnetic atoms.
\newblock {\em Science}, 352(6282):201--205, 2016.

\end{thebibliography}

\end{document}